%% file: main.tex
\documentclass[conference]{IEEEtran}
\IEEEoverridecommandlockouts
\pdfoutput=1
\def\BibTeX{{\rm B\kern-.05em{\sc i\kern-.025em b}\kern-.08em
    T\kern-.1667em\lower.7ex\hbox{E}\kern-.125emX}}

\IEEEoverridecommandlockouts
\usepackage{xspace}

\usepackage{cite}
\usepackage{amsmath,amssymb,amsfonts}
\usepackage{graphicx}
\usepackage{textcomp}
\usepackage{xcolor}

\usepackage{algpseudocode}
\usepackage{algorithm}
\usepackage{multirow}
\usepackage{makecell}
\usepackage[caption=false,font=footnotesize]{subfig}

\usepackage[bookmarks=true,breaklinks=true,letterpaper=true,colorlinks,citecolor=blue,linkcolor=blue,urlcolor=blue]{hyperref}

\graphicspath{{figs/}}

\begin{document}
\input{macro}
\title{\proj: Factor Graph Accelerator of LQR Control for Autonomous Machines}

\author{\IEEEauthorblockN{Yuhui Hao$^*$}
\IEEEauthorblockA{\textit{Tianjin University} \\
yuhuihao@tju.edu.cn}
\and
\IEEEauthorblockN{Bo Yu$^*$}
\IEEEauthorblockA{\textit{BeyonCa} \\
inzaghi1984@gmail.com}
\and
\IEEEauthorblockN{Qiang Liu$^\dagger$}
\IEEEauthorblockA{\textit{Tianjin University} \\
qiangliu@tju.edu.cn}
\and
\IEEEauthorblockN{Shao-Shan Liu}
\IEEEauthorblockA{\textit{PerceptIn} \\
shaoshan.liu@perceptin.io}
% \and
% \and
\thanks{$^*$ indicates equal contribution to the paper.}
\thanks{$\dagger$ indicates the corresponding author of the paper.}
}
\maketitle

\input{0.abstract.tex}
\input{1.introduction.tex}

\input{2.background.tex}
\input{3.hardware.tex}

\input{4.evaluation.tex}
\input{5.conclusion.tex}

\bibliographystyle{unsrt}
\bibliography{reference.bib}

\end{document}

%% file: macro.tex
%!TEX root=paper.tex

\newcommand{\website}[1]{{\tt #1}}
\newcommand{\program}[1]{{\tt #1}}
\newcommand{\benchmark}[1]{{\it #1}}
\newcommand{\fixme}[1]{{\textcolor{red}{\textit{#1}}}}
\newcommand{\answer}[1]{{\textcolor{blue}{\textit{#1}}}}

\newcommand*\circled[2]{\tikz[baseline=(char.base)]{
            \node[shape=circle,fill=black,inner sep=1pt] (char) {\textcolor{#1}{{\footnotesize #2}}};}}

\ifx\figurename\undefined \def\figurename{Figure}\fi
\renewcommand{\figurename}{Fig.}
\renewcommand{\paragraph}[1]{\textbf{#1} }
\newcommand{\figline}{{\vspace*{.05in}\hline}}

\newcommand{\Sect}[1]{Sec.~\ref{#1}}
\newcommand{\Fig}[1]{Fig.~\ref{#1}}
\newcommand{\Tbl}[1]{Tbl.~\ref{#1}}
\newcommand{\Equ}[1]{Equ.~\ref{#1}}
\newcommand{\Apx}[1]{Apdx.~\ref{#1}}
\newcommand{\Alg}[1]{Algo.~\ref{#1}}

\newcommand{\INPUT}{\item[\textbf{Input:}]}
\newcommand{\OUTPUT}{\item[\textbf{Output:}]}
\newcommand{\ALGORITHM}{\item[\textbf{Algorithm:}]}

\newcommand{\specialcell}[2][c]{\begin{tabular}[#1]{@{}c@{}}#2\end{tabular}}
\newcommand{\note}[1]{\textcolor{red}{#1}}

\newcommand{\proj}{\textsc{FGLQR}\xspace}
\newcommand{\mode}[1]{\underline{\textsc{#1}}\xspace}
\newcommand{\sys}[1]{\underline{\textsc{#1}}}

\newcommand{\no}[1]{#1}
\renewcommand{\no}[1]{}
\newcommand{\RNum}[1]{\uppercase\expandafter{\romannumeral #1\relax}}

\def\cA{{\mathcal{A}}}
\def\cF{{\mathcal{F}}}
\def\cN{{\mathcal{N}}}

\def\bA{{\mathbf{A}}}
\def\bB{{\mathbf{B}}}
\def\bb{{\mathbf{b}}}
\def\bC{{\mathbf{C}}}
\def\be{{\mathbf{e}}}
\def\bG{{\mathbf{G}}}
\def\bH{{\mathbf{H}}}
\def\bJ{{\mathbf{J}}}
\def\bK{{\mathbf{K}}}
\def\bM{{\mathbf{M}}}
\def\bMi{{\mathbf{M}^{-1}}}
\def\bN{{\mathbf{N}}}
\def\bP{{\mathbf{P}}}
\def\bQ{{\mathbf{Q}}}
\def\bp{{\mathbf{p}}}
\def\bR{{\mathbf{R}}}
\def\bS{{\mathbf{S}}}
\def\bSp{{\mathbf{S'}}}
\def\bU{{\mathbf{U}}}
\def\bUi{{\mathbf{U}^{-1}}}
\def\bV{{\mathbf{V}}}
\def\bW{{\mathbf{W}}}
\def\bX{{\mathbf{X}}}
\def\bs{{\mathbf{s}}}
\def\bnd{{\mathbf{n_d}}}
\def\bnm{{\mathbf{n_m}}}

\def\bnu{{\mathbf{n_u}}}
\def\bnr{{\mathbf{n_r}}}
\def\bnc{{\mathbf{n_c}}}

\def\DOF{\mathcal{DOF}}
\def\bzero{\mathbf{0}}
\def\bone{\mathbf{1}}

% checkmark and xmark in the pifont package
%\newcommand{\cmark}{\ding{51}}
%\newcommand{\xmark}{\ding{55}}

%% file: 0.abstract.tex
\begin{abstract}
 Factor graph represents the factorization of a probability distribution function and serves as an effective abstraction in various autonomous machine computing tasks. Control is one of the core applications in autonomous machine computing stacks. Among all control algorithms, Linear Quadratic Regulator (LQR) offers one of the best trade-offs between efficiency and accuracy. However, due to the inherent iterative process and extensive computation, it is a challenging task for the autonomous systems with real-time limits and energy constrained. 

 In this paper, we present \proj, an accelerator of LQR control for autonomous machines using the abstraction of a factor graph. By transforming the dynamic equation constraints into least squares constraints, the factor graph solving process is more hardware friendly and accelerated with almost no loss in accuracy. With a domain specific parallel solving pattern, \proj achieves $10.2\times$ speed up and $32.9\times$ energy reduction compared to the software implementation on an advanced Intel CPU. %Compared with the state-of-the-art accelerator that does not utilize factor graph, \proj achieves $2.4\times$ speedup.
 
\end{abstract}

\begin{IEEEkeywords}
factor graph, autonomous machine computing, computer architecture, robotics, optimal control, linear quadratic regulator
\end{IEEEkeywords}

%% file: 1.introduction.tex
\section{Introduction}
\label{sec::introduction}

Control is a critical component in many autonomous machine applications, such as autonomous vehicles, drones, and space exploration, as it allows robots to interact with and manipulate their environment in a precise and efficient way\cite{de2012theory}. An efficient controller enables the robot to execute actions precisely based on its objectives under the influence of many dynamic and uncertain factors. Without effective control, robots would be unable to perform complex tasks, navigate through complex environments, or respond to unexpected events. 

Optimal control is a branch of control theory that deals with finding the best possible control input for a given system to achieve a certain objective\cite{lewis2012optimal}.  In contrast to traditional control, which uses fixed and pre-determined control laws, optimal control finds the control input that minimizes a cost function, offering the best trade-offs between efficiency and accuracy. 

Linear Quadratic Regulator (LQR) is a popular optimal control algorithm used to control linear systems\cite{laine2019efficient}, which aims to minimize a quadratic cost function of the state error and control effort
% that represents the difference between the desired state and control of the system and the actual values, 
while enforcing the current state is determined by a linear function of the previous state and control. LQR provides an optimal control solution, reducing the amount of work done by the control systems engineer to tune the parameters. However, there are also limitations to the LQR controller. Traditional LQR requires the solution of the Riccati equation\cite{sideris2011riccati}, which is computationally intensive for large systems with high dimensions. This can lead to high power consumption, making it challenging to implement in real-time and embedded systems. 

Factor graphs have been used as efficient alternatives for solving optimization problems, especially in SLAM\cite{dellaert2021factor}. Factor graph is a bipartite graph that contains two types of nodes. Variable nodes represent the states to be estimated, and factor nodes indicate the relationships between states. Modeling the optimization problems as factor graphs makes them more intuitive and provides an efficient guide for solving them. 
% Although the LQR contains some equation constraints, by a sophisticated conversion, we can turn it into a Gaussian factor in a factor graph, and then solve the LQR problem with the powerful tool of factor graphs.
Some prior efforts have been made by using factor graph as the solver to LQR problem on CPU platform \cite{yang2021equality}. Compared with traditional solving methods, \cite{yang2021equality} speeds up the solving process to a certain extent, but still faces great challenges in implementing factor graph based LQR solver on resource-constrained edge devices, limiting its application in latency and power-sensitive applications. Unlike the SLAM problem, which is a least squares optimization problem, the LQR contains equation constraints formed by the dynamic equations of the system. The equation constraints pose additional computational burden in solving the optimization problem. There has been a recent proposal on building an LQR controller on an embedded CPU+FPGA SoC\cite{zhang2017design}. However, without factor graph as an abstraction, the performance of \cite{zhang2017design} could suffer less efficiency in handling large and complex LQR problems. To the best of our knowledge, there is no precedent for a factor graph accelerator that solves LQR problems.

In this paper, we propose a factor graph accelerator of LQR control for autonomous machines. By accelerating the LQR algorithmn using factor graph, we achieve $10.2\times$ speed up and $32.9\times$ energy reduction compared to an advanced CPU baseline. Our major contributions are:

%Compared with the state-of-the-art accelerator \cite{zhang2017design} that does not utilize factor graph, we achieve $2.4\times$ speedup. Our major contributions are:

\begin{itemize}
    \item By converting dynamic equation constraints of LQR into least squares constraints with large weights, the factor graph solving process is accelerated with almost no loss in accuracy.
    \item We propose a domain specific parallel solving pattern based on the structural properties of LQR factor graphs.
    \item We propose \proj, a hardware accelerator that exploits the potential parallel computation patterns to accelerate LQR algorithms.
\end{itemize}

The rest of this paper is organized as follows. \Sect{sec::background} introduces how to represent the LQR problem using factor graph and the solving process of factor graph. \Sect{sec::hardware} delves into the hardware design of the proposed \proj accelerator. \Sect{sec::evaluation} presents the evaluation results and we conclude in \Sect{sec::conclusion}.

\vspace{-10pt}

%% file: 2.background.tex
\section{LQR Factor Graph}
\label{sec::background}
In this section, we explain the background of LQR (\Sect{sec::bac::lqr}) and factor graph (\Sect{sec::bac::fg}), and describe how we model LQR using factor graph and how to solve it on the graph (\Sect{sec::bac::model}).

% \vspace{-5pt}
\subsection{LQR}
\label{sec::bac::lqr}
% LQR is a state feedback controller that solves the optimal controls for a linear system with quadratic costs on control effort and state error. We consider the finite-horizon, discrete LQR problem. The task is to find the optimal controls $u_k$ at time instances $k$ so that a total cost is minimized. The LQR problem can be represented as a constrained optimization problem where the costs of control and state error are represented by the minimization objective as \Equ{equ::lqr}, and the system dynamics are represented by the constraints as \Equ{equ::dyn}. $x_k$ and $u_k$ are the state and control. $Q$ and $R$ are weight matrices.

LQR is a state feedback controller that optimally computes controls for a linear system with quadratic costs on control effort and state error~\cite{fglqr}. In this work, we focus on the finite-horizon, discrete LQR problem, aiming to determine the optimal controls $u_k$ at time instances $k$ to minimize the total cost. The LQR problem can be formulated as a constrained optimization problem, where the control and state error costs are represented as the minimization objective (\Equ{equ::lqr}), and the system dynamics are represented as constraints (\Equ{equ::dyn}). Here, $x_k$ and $u_k$ represent the state and control variables, respectively, while $Q$ and $R$ denote the weight matrices.

\begin{small}
\begin{equation}
    \mathop{\arg\min}\limits_{u_0,\dots,u_{N-1}} \{x_N^TQx_N+\sum_{k=0}^{N-1}x_k^TQx_k+u_k^TRu_k\}
    \label{equ::lqr}
\end{equation}
\begin{equation}
    s.t.\ x_{k+1}=Ax_k+Bu_k
    \label{equ::dyn}
\end{equation}
\end{small}

\vspace{-5pt}
\subsection{Factor Graph}
\begin{figure}[t]
    \centering
    \includegraphics[trim=0 0 0 0, clip, width=0.7\columnwidth]{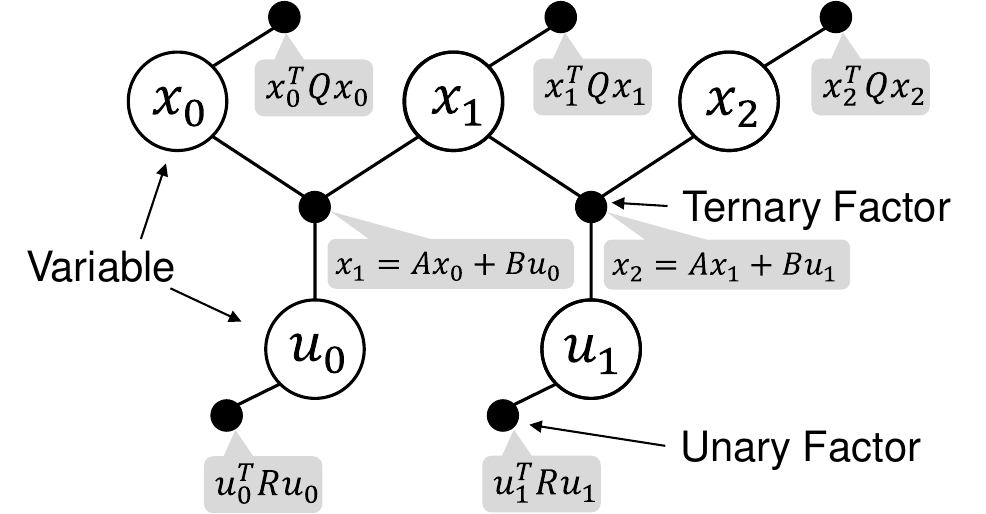}
    \caption{LQR factor graph with time horizen $N=2$. The variables represent the states and controls. The ternary factors represent the dynamics model constraints and the unary factors represent the state and control costs.}
    \label{fig::lqrfg}
    \vspace{-15pt}
\end{figure}

\label{sec::bac::fg}
Factor graph is a bipartite graph consisting of variables connected by factors, where a factor can be viewed as a least squares objective over the variables it is connected to. The factor nodes capture the relationships between the variables, providing a compact and efficient representation of the problem.

Factor graphs are a powerful tool in solving least squares problems as \Equ{equ::ls}, where $\Vert\cdot\Vert^2_{\Sigma_k}=(\cdot)^T\Sigma_k^{-1}(\cdot)$ is the Mahalanobis norm. $\bA$ contains all $\Sigma_k^{-\frac{1}{2}}A_k$ and $\bb$ stacks all $\Sigma_k^{-\frac{1}{2}}b_k$ vertically. $\Sigma_k$ is the covariance matrix indicating the weight of the item in the overall. The smaller the covariance, the greater the weight of the item. 
% where $\bA$ contains all $\Sigma_k^{-\frac{1}{2}}A_k$ and $\bb$ stacks all $\Sigma_k^{-\frac{1}{2}}b_k$ vertically. $\Vert\cdot\Vert^2_{\Sigma_k}=(\cdot)^T\Sigma_k^{-1}(\cdot)$ is the Mahalanobis norm, where $\Sigma_k$ is the covariance matrix indicating the weight of the item in the overall. The smaller the covariance, the greater the weight of the item. 
\begin{equation}
    \mathop{\arg\min}\limits_\bX\sum_k \Vert A_kX_k-b_k\Vert_{\Sigma_k}^2=\mathop{\arg\min}\limits_\bX\Vert \bA\bX-\bb\Vert^2
    \label{equ::ls}
\end{equation}

Solving \Equ{equ::ls} is equivalent to solving the system of linear equations $\bA\bX = \bb$. This involves the QR decomposition on matrix $\bA$, which decomposes $\bA$ into the product of an orthogonal matrix  and an upper triangular matrix $\bR$. However, decomposing the matrix directly requires significant costs on latency and memory. Since the structure of the factor graph is equivalent to the sparsity pattern of matrix $\bA$, factor graph helps us to solve it in an incremental way, reducing the time and space complexity. 

\vspace{-5pt}
\subsection{Modeling and Solving}

\begin{figure}[t]
    \centering
    \includegraphics[trim=0 0 0 0, clip, width=0.85\columnwidth]{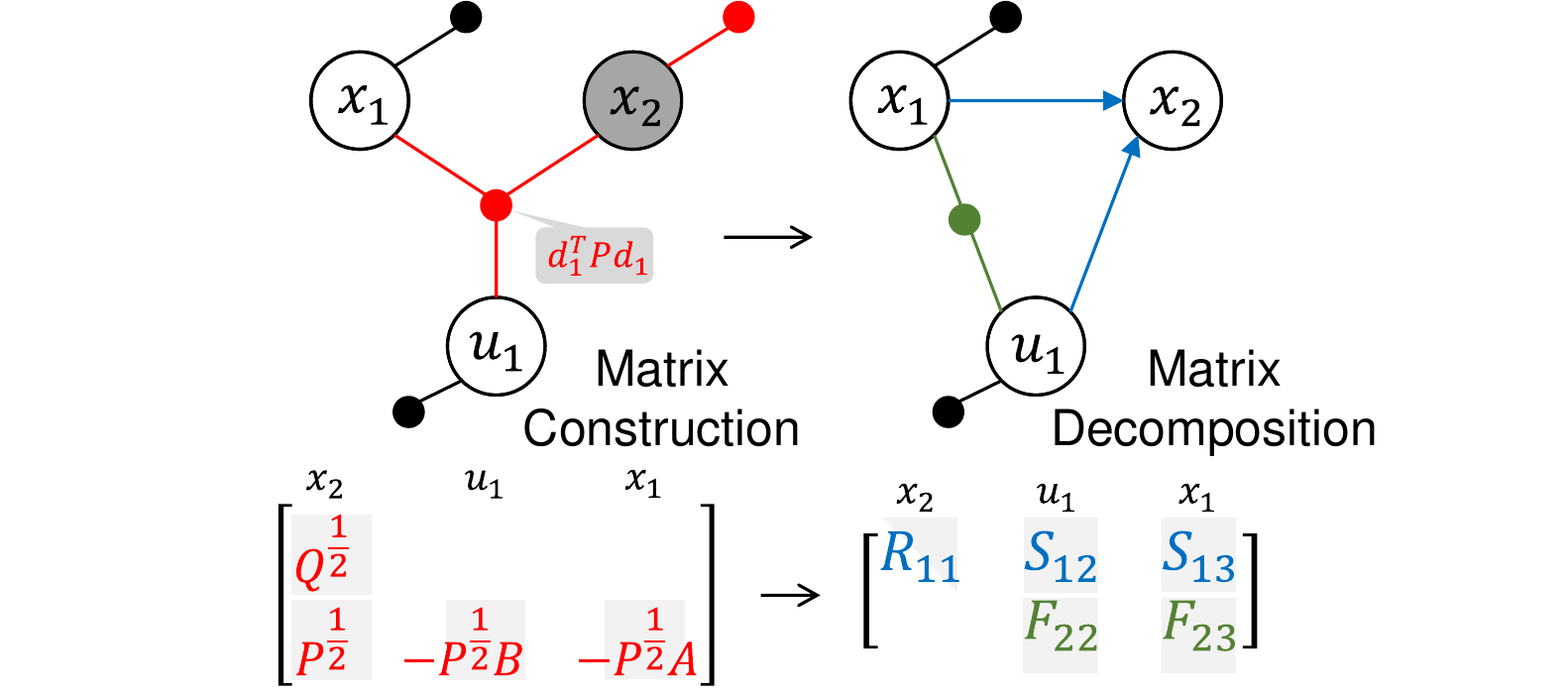}
    \caption{The process of solving the LQR factor graph. The figure shows the matrix operations involved and the changes in the factor graph when eliminating variable $x_2$. The shading in the matrix shows the shape of the non-zero matrix blocks.}
    \label{fig::ve}
    \vspace{-10pt}
\end{figure}

\label{sec::bac::model}
\paragraph{Modeling.} \Fig{fig::lqrfg} shows the LQR factor graph with the oldest states and controls on the left, and the newest states and controls on the right. The ternary factors represent the dynamics model constraints, and the unary factors represent the state and control costs we seek to minimize via least-squares.

\paragraph{Solving.} To solve the LQR problem efficiently using factor graphs, we propose to transform \Equ{equ::lqr} and \Equ{equ::dyn} into a least-squares form without equation constraints, as \Equ{equ::lslqr}, and increase the weight of the dynamic constraints $P$. In this way, all the factor types in the factor graph are all the same, which facilitates the computation and hardware design.
\begin{small}
\begin{equation}
\begin{aligned}
    \mathop{\arg\min}\limits_{u_0,\dots,u_{N-1}} \{x_N^TQx_N&+\sum_{k=0}^{N-1}x_k^TQx_k+u_k^TRu_k+ d_k^TPd_k\} \\
                                                         d_k &= x_{k+1}-(Ax_k+Bu_k)
    \label{equ::lslqr}
\end{aligned}
\end{equation} 
\end{small}

% 正常的消元顺序
With the redefinition of dynamic factors, solving factor graphs becomes more efficient and concise. The process of solving the LQR factor graph is eliminating each variable, which is divided into two steps, as shown in \Fig{fig::ve}. \textbf{Matrix Construction.} Given a specific order of variable nodes, when eliminating a variable, the factor nodes adjacent to it are selected. The least-squares terms represented by these factors form a small matrix (Red part). \textbf{Matrix Decomposition.} Perform a partial QR decomposition on this small matrix. Transform the column corresponding to this variable into the upper triangular form. The first block row of the decomposed matrix is used as part of the upper triangular matrix $\bR$ (Blue part), and the second block row is rejoined to the factor graph as a new factor (Green part). These steps are done incrementally until all variables are eliminated. Finally, we obtain the complete upper triangular matrix $\bR$.

% 并行消元
\paragraph{Parallel Solving.} According to the traditional LQR algorithm\cite{sideris2011riccati}, the control is a linear function of the current state, \textit{i.e.,} $u_k = K_kx_k$. Therefore, it must be eliminated sequentially from the new time steps to the old to calculate the gain matrix $K$ and thus $u$. However, since all we require is a series of values of $u_k$ to make $x_k$ converge quickly, it is possible to start elimination from multiple variables without shared factors simultaneously on the factor graph. It is the same solution as the factor graph with sequential elimination, but significantly accelerates the computation. We propose an elimination order applicable to LQR factor graphs, eliminating variables from both sides of the graph towards the middle, respectively. Since there is no shared factor when eliminating variables on both sides, it can be executed in hardware in parallel, which accelerates the solving process.

%% file: 3.hardware.tex
\section{Hardware Architecture}
\label{sec::hardware}

In this section, we present an overview of the proposed \proj accelerator (\Sect{sec::hw::overview}) as well as the detailed implementation of each sub-block (\Sect{sec::hw::whitening} to \Sect{sec::hw::bs}). Several leveraged optimization techniques are provided.

\vspace{-5pt}

\subsection{Hardware Architecture Overview}
\label{sec::hw::overview}
\begin{figure}[t]
\setlength{\abovecaptionskip}{0.cm}
    \centering
    \includegraphics[trim=0 0 0 0, clip, width=\columnwidth]{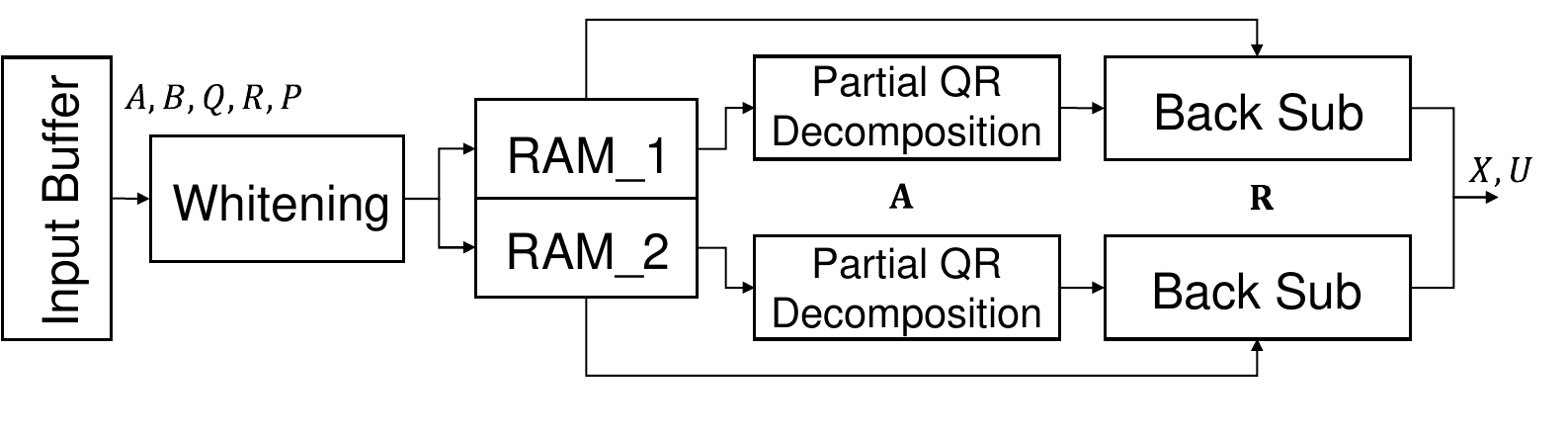}
    \caption{Overview of the proposed \proj accelerator.}
    \label{fig::hw}
    \vspace{-15pt}
\end{figure}

\Fig{fig::hw} shows the \proj hardware architecture of the accelerator. The \textbf{Matrix Construction} as well as the \textbf{Matrix Solving} process is performed on the accelerator. The whitening Block completes the construction of the coefficient matrix and right-hand side vector. Matrix solving includes matrix decomposition and back substitution. Since the LQR factor graph can be eliminated in parallel, we design two sets of matrix computation and storage units to accelerate the whole solving process.

% \vspace{-5pt}

\subsection{Whitening Block}
\label{sec::hw::whitening}
The whitening block performs the construction of the coefficient matrix $\bA$, which requires the multiplication of the weight matrix $P^{\frac{1}{2}}$ with the state matrix $A$ and the input matrix $B$, respectively. In fact, $P$ is usually a diagonal matrix, so this matrix multiplication can be optimized. Since the diagonal matrix only contains non-zero values along the main diagonal, the same value is multiplied by the corresponding row of $A$ and $B$. It allows performing all the multiplications in parallel for a given row.

In addition, since $P$ is the weight of the dynamic least square  constraint which replaces the equation constraint, a large value of $P$ is required. We set the main diagonal element value to an integer power of 2, e.g., $2^{10}$. This allows exponential bit addition instead of floating point multiplication to speed up the operation, which achieves $6.35\times$ speedup in our design.

\begin{figure}[t]
\setlength{\abovecaptionskip}{0.cm}
\centering
    \begin{minipage}{0.46\columnwidth}
    \centering
    \includegraphics[trim=0 0 0 0, clip, width=\columnwidth]{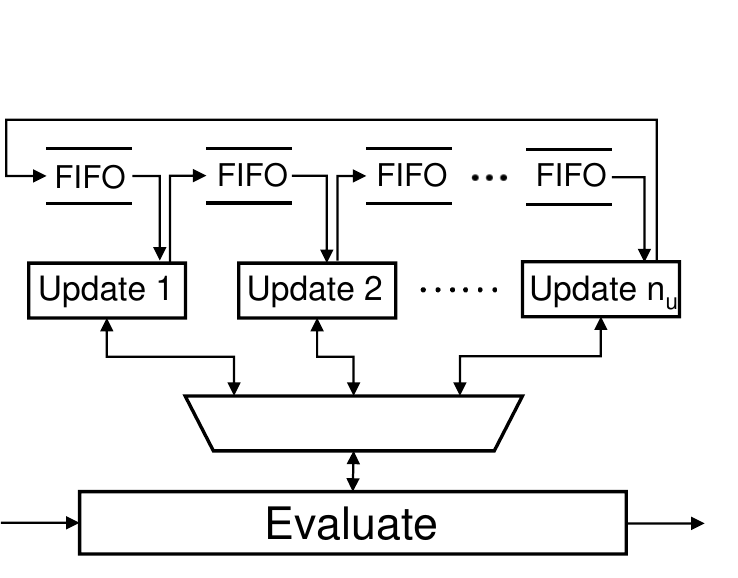}
    \caption{The partial QR decomposition block uses one Evaluate unit and $\bnu$ time-multiplexed Update units to ensure a balanced pipeline.}
    \label{fig::qr}
    \end{minipage}
    \hspace{2pt}
    \begin{minipage}{0.44\columnwidth}
    \centering
    \includegraphics[trim=0 0 0 0, clip, width=\columnwidth]{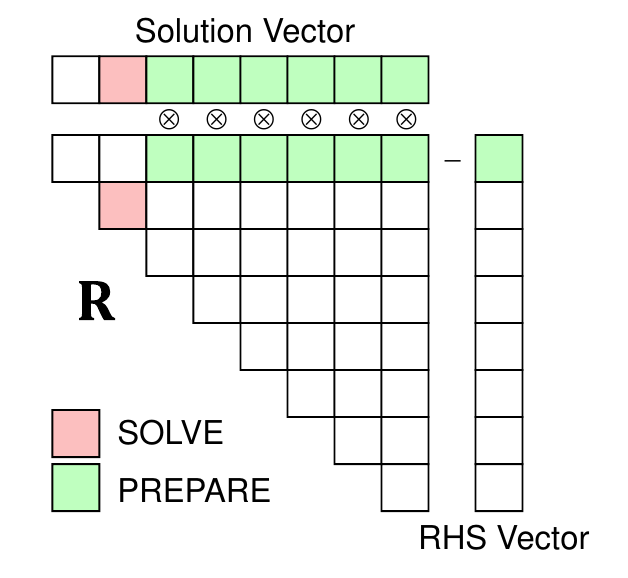}
    \caption{The operation involved in the PREPARE stage (green) and the SOLVE stage (red).}
    \label{fig::bacsub}
    \end{minipage}
    \vspace{-15pt}
\end{figure}

\subsection{Partial QR Decomposition Block}
\label{sec::hw::qr}

% The partial QR decomposition block applies the Householder transformation for each column of the input matrix to make the corresponding elements below the diagonal equal to zero. It needs two phases. In the Evaluate phase, the Householder matrix $\bH$ is constructed from the column to be zeroing. In the Update phase, the entries below the diagonal of this column are set to zeros by left multiplying $\bH$, and the following columns are updated. The updated matrix in the lower right serves as the input for the next iteration. The iteration continues until the first variable column is triangulated.

The partial QR decomposition block employs Householder transformations to eliminate elements below the diagonal in the input matrix, making them zero. This process consists of two phases. In the Evaluate phase, a Householder matrix $\bH$ is constructed based on the column to be zeroed. In the Update phase, the entries below the diagonal of this column are transformed to zeros by performing a left multiplication with $\bH$, and subsequent columns are updated accordingly. The resulting updated matrix in the lower-right portion becomes the input for the next iteration. This iterative process continues until the first variable column is triangulated.

% \Fig{fig::qr} shows the architecture of this block. Based on our analysis of the data dependencies, the Evaluate-Update phase can be pipelined. The current iteration of the Update phase and the next iteration of the Evaluate phase have no data dependency and thus can run in parallel. As the Update phase is the bottleneck in the pipeline, we design $\bnu$ number of time-multiplexed Update units, each of which is connected through a FIFO due to the sequential data read/write relationship. The performance improves and converges as we increase the $\bnu$, alongside the increase of the cost of hardware resources.

\Fig{fig::qr} depicts the architecture of this block. Through data dependency analysis, we find that the Evaluate-Update phase can be pipelined. The current iteration of the Update phase and the next iteration of the Evaluate phase have no data dependency, allowing them to execute in parallel. As the Update phase is the bottleneck in the pipeline, we design $\bnu$ time-multiplexed Update units, each connected via a FIFO to accommodate the sequential data read/write operations to enhance performance. As we increase $\bnu$, the performance improve, but this comes with a corresponding increase in hardware resource consumption.
% All of the Update units are connected to the Evaluate unit. 
% between the front and back units

\vspace{-5pt}

\subsection{Back Substitution Block}
\label{sec::hw::bs}
The back substitution block receives as input the non-zero values in the upper triangular matrix $\bR$ and the right-hand side vector $\bb$. It performs back substitution to solve for the states and controls.

Back substitution involves a nested loop process so we pipelined to speed up the computation. We divide it into four stages, including FETCH, PREPARE, SOLVE, and WRITE. Due to space limitations, we describe the PREPARE and EXECUTE phases specifically.
\Fig{fig::bacsub} shows the operation involved in the two stages. The PREPARE stage prepares the data for the SOLVE stage. In this stage, it calculates the right-hand side of the equation by subtracting the products of the upper triangular elements and the solution vector from the right-hand side vector. The SOLVE stage calculates the solution for the current row by dividing the result of the PREPARE stage by the diagonal element of the matrix.

%% file: 4.evaluation.tex
\section{Evaluation Results}
\label{sec::evaluation}

This section evaluates the proposed \proj accelerator through a series of experiments. \Sect{sec::sec::setup} introduces the experimental setup. \Sect{sec::sec::accuracy} evaluates the control accuracy of the accelerator. \Sect{sec::sec::performance} demonstrates the speedup and energy reduction of the proposed accelerator compared to software implementation and state-of-the-art accelerators. \Sect{sec::sec::resource} gives the hardware resource utilization.

\subsection{Experimental Setup}
\label{sec::sec::setup}
\paragraph{Hardware Platform.} We develop and synthesize the proposed \proj accelerator using Vitis-HLS tool on a Xilinx Zynq-7000 SoC ZC706 FPGA. The accelerator operates at a fixed frequency of 167 MHz. The Vivado power analysis tool is used to evaluate the power consumption of the accelerator under different test workloads.

\paragraph{Baselines.}
A software implementation of LQR controller, PathTracking\cite{tracking}, is used as a software baseline, which uses GTSAM\cite{gtsam} to implement factor graph optimization. The software is evaluated on the 11th Intel processor operating at 2.5 GHz.

\paragraph{Experimental Scenario.} The proposed \proj accelerator controls an autonomous vehicle to follow a pre-planned trajectory in a two-dimensional plane. The LQR problem is given by \Equ{equ::lslqr}, where $x=[d,\dot d,\theta, \dot\theta,v]^T\in\mathbb{R}^5$, $u=[\gamma,a]^T\in\mathbb{R}^2$, $Q=I_{5\times5}$, $R=I_{2\times2}$, $P=2^{10}\cdot I_{5\times5}$. $d$, $\theta$, and $v$ are the distance, angle, and speed deviation from the reference value, respectively. $\dot d$ and $\dot\theta$ are the derivatives of $d$ and $\theta$. $\gamma$ and $a$ are the steering angle and acceleration of the vehicle. 
\vspace{-5pt}

\subsection{Control Accuracy}
\label{sec::sec::accuracy}
We evaluate the accuracy of the proposed \proj accelerator, which is measured by the distance deviation from the reference trajectory. The accumulated sum of squares of the Euclidean distance between the actual position of each time step and its target position is used as the cost value. The smaller the cost value, the more accurate it is.

\begin{table}[h]
    \vspace{-5pt}
    \centering
    \caption{The cost value for the four variants. The smaller the cost value, the more accurate it is.}
    \label{tab::accuracy}
    \begin{tabular}{|c|c|c|c|c|}% 通过添加 | 来表示是否需要绘制竖线
    \hline  % 在表格最上方绘制横线
     Variants  & \textbf{Trad} & \textbf{EquCons} & \textbf{Sequ} &\textbf{Ours}\\
    \hline  %在第一行和第二行之间绘制横线
    Cost & 48.21 & 48.63 & 48.74  & 48.74 \\
    \hline % 在表格最下方绘制横线
    \end{tabular}
    
    \vspace{-5pt}
\end{table}

We test with four variants. \textbf{Trad} denotes solving the LQR problem using the traditional Riccati equation. \textbf{EquCons} represents solving the LQR problem using factor graph subject to the dynamic equation constraint. \textbf{Sequ} indicates solving the LQR problem using factor graph with dynamic least squares constraints and eliminating variables in sequential. \textbf{Ours} is the result of the proposed \proj accelerator, which solves the LQR problem using factor graph with dynamic least squares constraints and eliminating variables in parallel. All variants use single-precision floating-point numbers.

\Tbl{tab::accuracy} shows the results of the cost value for the four variants. The results show that the parallel elimination mode in \textbf{Ours} does not affect the accuracy. However, it can execute in parallel on hardware implementation to accelerate the solving process. The transformation of the dynamic equation constraint into least squares constraint affects the accuracy negligibly, but facilitates using factor graphs to solve the LQR problem. 
% In fact, on the software side, this transformation also accelerates the computation, which we do not overstate here since we focus on hardware acceleration.

\subsection{Performance Evaluation}
\label{sec::sec::performance}

We evaluate the runtime latency and energy efficiency of the proposed \proj accelerator compared with the CPU baseline. Both latency and energy consumption are obtained from solving the LQR problem with time horizen $N=50$.

%\Fig{fig::speedup} shows the runtime improvements over two baselines. 

\Tbl{tab::perf} demonstrates that %\proj significantly reduces the control latency. 
compared to the CPU baseline, \proj achieves an average speed up of $10.2\times$. The average latency of one iteration in LQR control has been reduced to $1.94$ ms. Compared to the CPU baseline, \proj hardware saves the energy by $32.9\times$. 

\begin{table}[h]
\vspace{-5pt}
    \centering
    \caption{Speedup and energy reduction achieved by \proj compared to the Intel CPU baseline.}
    \label{tab::perf}
    \begin{tabular}{|c|c|c|}% 通过添加 | 来表示是否需要绘制竖线
    \hline  % 在表格最上方绘制横线
          & \textbf{Speedup ($\times$)} & \textbf{Energy Reduction ($\times$)}\\
    \hline  %在第一行和第二行之间绘制横线
     \proj & 10.2 & 32.9\\
    \hline % 在表格最下方绘制横线
    \end{tabular}
    
    \vspace{-5pt}
\end{table}

%\Fig{fig::energy} demonstrates the reduction of energy consumption. Compared to the CPU baseline, \proj hardware saves the energy by $32.9\times$.

\subsection{Resource Utilization}
\label{sec::sec::resource}

\Tbl{tab::resource} demonstrates the resource utilization of the proposed \proj accelerator under our experimental scenario. The FPGA device has 218K LUTs, 437K Flip-Flops, 545 BRAMs and 900 DSPs in total. Overall, the hardware architecture consumes 45\% LUTs, 29\% Flip-Flops, 15\% BRAMs and 51\% DSPs.

\begin{table}[h]
    \vspace{-5pt}
    \centering
    \caption{Resource utilization (utilization percentages and absolute numbers) of \proj accelerator.}
    \label{tab::resource}
    \begin{tabular}{|c|c|c|c|c|}% 通过添加 | 来表示是否需要绘制竖线
    \hline  % 在表格最上方绘制横线
      Resource & \textbf{LUT} & \textbf{FF} & \textbf{BRAM} & \textbf{DSP}\\
    \hline  %在第一行和第二行之间绘制横线
    \proj & \specialcell{45\%\\(98641)} & \specialcell{29\%\\(127460)} & \specialcell{15\%\\(84)}  & \specialcell{51\%\\(464)} \\
    \hline % 在表格最下方绘制横线
    \end{tabular}
    
\end{table}

\vspace{-10pt}

%% file: 5.conclusion.tex
\section{Conclusion}
\label{sec::conclusion}

This paper takes the first step of utilizing the factor graph as an abstraction to build an accelerator for control algorithms in autonomous machine applications. We propose, \proj, a factor graph accelerator of LQR control for autonomous machines. By transforming the dynamic equation constraints into least squares constraints, the LQR algorithm is reformulated and the factor graph solving process is more hardware friendly and accelerated with almost no loss in accuracy. With a domain specific parallel solving pattern, \proj achieves $10.2\times$ speed up and $32.9\times$ energy reduction compared to the software implementation on an advanced Intel CPU. %Compared with the state-of-the-art accelerator that does not utilize factor graph, \proj achieves $2.4\times$ speedup.